\documentstyle{article}
\newcommand{\gf}{\mbox{$\gamma_{5}$}}

\newcommand{\one}{\mbox{${\bf 1}$}}

\newcommand{\be}{\begin{equation}}
\newcommand{\ee}{\end{equation}}
\newcommand{\bea}{\begin{eqnarray}}
\newcommand{\eea}{\end{eqnarray}}
\newcommand{\beas}{\begin{eqnarray*}}
\newcommand{\eeas}{\end{eqnarray*}}
\newcommand{\bdm}{\begin{displaymath}}
\newcommand{\edm}{\end{displaymath}}
\newcommand{\ba}{\begin{array}}
\newcommand{\ea}{\end{array}}
\newcommand{\bi}{\begin{itemize}}
\newcommand{\ei}{\end{itemize}}
\newcommand{\ben}{\begin{enumerate}}
\newcommand{\een}{\end{enumerate}}
\newcommand{\bc}{\begin{center}}
\newcommand{\ec}{\end{center}}
\newcommand{\bfl}{\begin{flushleft}}
\newcommand{\efl}{\end{flushleft}}
\newcommand{\bfr}{\begin{flushright}}
\newcommand{\efr}{\end{flushright}}
\newcommand{\bd}{\begin{description}}
\newcommand{\ed}{\end{description}}
\newcommand{\bq}{\begin{quote}}
\newcommand{\eq}{\end{quote}}
\newcommand{\bfg}{\begin{figure}}
\newcommand{\efg}{\end{figure}}
\newcommand{\bt}{\begin{table}}
\newcommand{\et}{\end{table}}
\newcommand{\btb}{\begin{tabular}}
\newcommand{\etb}{\end{tabular}}
\newcommand{\btg}{\begin{tabbing}}
\newcommand{\etg}{\end{tabbing}}

\newcommand{\kslash}
           {\mbox{$ k \hspace{-1.1ex} \mbox{/} \hspace{-0.07ex} $}}
\newcommand{\lslash}
           {\mbox{$ l \hspace{-0.9ex} \mbox{/} \hspace{-0.15ex} $}}

\newcommand{\qslash}
           {\mbox{$ q \hspace{-1.1ex} \mbox{/} \hspace{-0.05ex} $}}

\relax
\def\frac#1#2{{#1\over#2}}
\def\<{\langle}\def\>{\rangle}
\newcount\parenthesis \parenthesis=0 \newcount\n
\def\({\global\advance\parenthesis by1\left(}
\def\){\global\advance\parenthesis by-1\right)}
\def\{{\global\advance\parenthesis by1\left\lbrace}
\def\}{\global\advance\parenthesis by-1\right\rbrace}
\def\[{\relax} % dummy parenthesis
\def\]{\relax} % dummy parenthesis
\def\triloop#1\repeat{\global\n=0\global\let\body=#1\iterate}
\def\iterate{\body\let\next=\iterate\else\let\next=\relax\fi\next}
\def\ldd{\ifnum\n<\parenthesis\global\advance\n by1
\left.\nulldelimiterspace=0pt\mathsurround=0pt}
\def\rdd{\ifnum\n<\parenthesis\global\advance\n by1
\right.\nulldelimiterspace=0pt\mathsurround=0pt}
\def\nl{\triloop\rdd\repeat\hfill\cr\qdd\triloop\ldd\repeat{}}

\def\qdd{\quad\quad}
\newcount\exacount\exacount=0\font\caps=cmcsc10
\def\Istrut{\vrule height11pt depth4pt width0pt}
\def\TRIexa#1#2#3#4{\global\advance\exacount by1\par\filbreak
{\offinterlineskip
  \vbox{\hrule\hbox to\hsize{\Istrut\vrule
      \hbox to 8mm{\hfil\caps\the\exacount\hfil}\vrule
      \quad\rm#1\hfill\vrule
      \hbox to 32mm{\hfill{\caps Mode: }{\tt #2}\hfill}\vrule
      \hbox to 32mm{\hfill{\caps Tolerance: }{\tt #3}\hfill}\vrule}
    \hrule\hbox to\hsize{\Istrut\vrule\hfill#4\hfill\vrule}\hrule}
}\nobreak}
\begin{document}
\title{\raisebox{2cm}{\makebox[0pt][r]{\small UTAS-PHYS-93-40}} Tensor
Integrals for Two Loop Standard Model Calculations}
\author{Dirk Kreimer\thanks{email: kreimer@physvax.phys.utas.edu.au}\\
\small Dept.~of Physics\\
\small Univ.~of Tasmania\\
\small G.P.O.Box 252C\\
\small Hobart, 7001\\
\small Australia}
\date{December 1993}
\maketitle
\begin{abstract}
We give a new method for the reduction of tensor integrals to finite
integral representations and UV divergent analytic expressions.
This includes a new method for the handling of the $\gamma$-algebra.
\end{abstract}
In calculating two loop corrections to the Standard Model one is confronted
with two main problems. One is the analytical difficulty
of integrals involving different masses. Often one is restricted to
approximations as for example zero mass assumptions for light particles
\cite{1}.
A further problem stems from the difficulty in the algebraic sector,
where the increasing number
of terms for the tensor structure has to be handled. For the case of
two-point
functions an elegant method is available to reduce all tensor integrals
to a basic set of scalar integrals \cite{2}. Nevertheless it is clear
that these methods will run into difficulties when applied to three-
or four-point functions. Also, for arbitrary mass cases, one is restricted to
either
numerical evaluations or asymptotic expansions \cite{bas}. This is due to the
fact that
for
arbitrary mass cases it is not possible to express the generic scalar
integrals through known special functions \cite{3}.

In this paper we will present a general method to express two-loop integrals
in terms of finite integrals suitable to numerical evaluations plus a set
of products of one-loop integrals containing the UV singular part.
The method presented here is as well applicable to arbitrary tensor integrals
but will prove especially powerful when applied to two-loop graphs
directly, as it will bypass the whole tensor structure problem by
constructing what can be called the characteristic polynomial
of the graph.

The final result is appropriate for the use of integral representation
generated from the ones in \cite{dirk} but the method can also be applied
with other choices for the generic integrals, as the way how we
handle the tensor structure is independent from the analytical approach one
uses.

The methods presented in the following apply to two-loop $n$-point
Green's function for arbitrary $n$, but in examples we will
restrict ourself to two-loop two- and three-point problems.
We will demonstrate the use of our method on the
topology of the following figure.\\[5mm]
%_________________________________________________________________________
% FIGURE
%
\unitlength=1.00mm
%\special{em:linewidth 0.4pt}
\linethickness{0.4pt}
\begin{picture}(33.00,14.00)
\put(16.00,7.00){\circle{14.00}}
\put(23.00,7.00){\line(1,0){10.00}}
\put(9.00,7.00){\line(-1,0){9.00}}
\put(16.00,13.00){\line(0,-1){13.00}}
\put(16.00,13.00){\line(0,1){1.00}}
\put(9.00,14.00){\makebox(0,0)[cc]{1}}
\put(23.00,14.00){\makebox(0,0)[cc]{5}}
\put(9.00,0.00){\makebox(0,0)[cc]{2}}
\put(23.00,0.00){\makebox(0,0)[cc]{4}}
\put(18.00,7.00){\makebox(0,0)[cc]{3}}
\end{picture}\nopagebreak
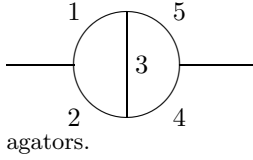
{\small Fig.1: The master two-loop
two-point function; the labels denote the propagators.\\[5mm]}

A general two-loop integral has the following  representation
\bea
I=\int d^D ld^D k \frac{T^{(n_l,n_k)}}{N_l M N_k},\label{e1}
\eea
where $(n_l,n_k)$ denotes the rank of its tensor structure in $l,k$ loop
momenta. Here $N_l$ is the part of the denominator containing only
propagators depending on $l$, $N_k$ the same for loop momentum $k$ and
$M$ contains the propagators involving both loop momenta.

We omit in the following the cases where terms in the numerator cancel
the $M$ part of the denominator completely out as these terms give rise
to products of one-loop integrals only.

Let us assume $I$ has an overall degree of divergence $\omega$ and degrees of
divergence
$\omega_l,\omega_k$ for the subdivergences. To make this integral convergent
we have
to subtract its $l$ resp.~$k$ subdivergent behaviour as well as its overall
divergence.

To this end let us define the following
\bea
\tilde{N_l} & = & N_l\mid_{m_i=0,q_i=0},\nonumber\\
\tilde{N_k} & = & N_k\mid_{m_i=0,q_i=0},\nonumber\\
\tilde{M} & = & M\mid_{m_i=0,q_i=0},\label{e2}\\
L^{j_l} & = & \frac{(\tilde{N_l}\tilde{M}-N_l
M)^{j_l}}{(\tilde{N_l}\tilde{M})^{j_l}},
 \nonumber\\
K^{j_k} & = & \frac{(\tilde{N_k}\tilde{M}-N_k
M)^{j_k}}{(\tilde{N_k}\tilde{M})^{j_k}},
\nonumber
\eea
where $\mid_{m_i =0,q_i =0}$ means to evaluate the corresponding expression
with all masses and external momenta set to zero.

Now consider the replacement
\beas
\frac{1}{N_l} \rightarrow \frac{1}{N_l}L^{j_l},
\eeas
which improves $\omega$ and $\omega_l$ at least by $j_l$ and leaves $\omega_k$
unchanged, and a similar replacement for $k$.
Consequently
\bea
\frac{1}{N_l M N_k} \rightarrow\frac{1}{N_l M N_k} K^{j_k} L^{j_l},\label{e3}
\eea
results (at least) in an improvement
\beas
\omega \rightarrow \omega+j_l+j_k =: \omega^\prime,\\
\omega_l \rightarrow \omega_l+j_l =:\omega_l^\prime,\\
\omega_k \rightarrow \omega_k+j_k =:\omega_k^\prime.
\eeas
Let us introduce the notation ${\underline \omega^\prime>0} \Leftrightarrow
\{\omega^\prime >0,\omega_l^\prime >0,\omega_k^\prime >0\}$.
Let us further define the notation $C_0:=C-1$ for arbitrary expressions
$C$ so that
\bea
L^{j_l}_0 & = &
   \frac{1}{(\tilde{N_l}\tilde{M})^{j_l}}\left[
   \sum_{l=1}^{j_l}{j_l \choose l}(-N_l M)^l (\tilde{N_l}\tilde{M})^{j_l-l}
   \right],\nonumber\\
K^{j_k}_0 & = & \frac{1}{(\tilde{N_k}\tilde{M})^{j_k}}\left[
   \sum_{k=1}^{j_k}{j_k \choose k}(-N_k M)^k (\tilde{N_k}\tilde{M})^{j_k-l}
   \right],\nonumber\\
   \left[ L^{j_l}K^{j_k}\right] {}_0 & = &
   \frac{1}{(\tilde{N_l}\tilde{M})^{j_l}}
   \frac{1}{(\tilde{N_k}\tilde{M})^{j_k}}
   \left[
   \sum_{l=0}^{j_l}
   \sum_{\stackrel{k=0}{l+k>0}}^{j_k}\right.\nonumber\\
   & & \left. {j_l \choose l}{j_k \choose k}(-N_l M)^l
(\tilde{N_l}\tilde{M})^{j_l-l}
   (-N_k M)^k (\tilde{N_k}\tilde{M})^{j_k-k}
   \right]
   \label{e4}
\eea
where we emphasize that the sums start with $l=1$ resp.~$k=1$.

We then have the algebraic identity
\bea
\frac{T^{(n_l,n_k)}}{N_l M N_k} & = &
\frac{T^{(n_l,n_k)}}{N_l M N_k} \left[ L^{j_l} K^{j_k}\right]-
  \frac{T^{(n_l,n_k)}}{N_l M N_k}\left[
  L^{j_l} K^{j_k}\right]_0 \;\;\forall j_l,j_k.\label{e5}
\eea

Now there clearly exists for a given $T$ and given $N_l, M,N_k$
powers $j_l,j_k$ such that $\omega^\prime, \omega_l^\prime ,
\omega_k^\prime $ are all positive, $\underline{\omega}^\prime>0$, so that
the first expression on
the right hand side of equation (\ref{e5}) is finite. So it can be calculated
in
$D=4$ dimensions.
But by inspecting the sums
\beas
L^{j_l} & = & \frac{1}{(\tilde{N_l}\tilde{M})^{j_l}}\left[
   \sum_{l=0}^{j_l}{j_l \choose l}(-N_l M)^l (\tilde{N_l}\tilde{M})^{j_l-l}
   \right],\\
K^{j_k} & = & \frac{1}{(\tilde{N_k}\tilde{M})^{j_k}}\left[
   \sum_{k=0}^{j_k}{j_k \choose k}(-N_k M)^k (\tilde{N_k}\tilde{M})^{j_k-l}
   \right],
\eeas
we can identify all terms in the expansion which fulfil
$\underline{\omega}>0$. As these terms can be calculated in four dimensions
they cancel against the corresponding terms of the second expression
in the right hand side of equation (\ref{e5}).
Note that to identify the divergent part a simple
 counting of dimensions
 of masses and exterior momenta in the numerator of equation (\ref{e5})
 is sufficient.

So we define the following expressions
\bea
\hat{L^{j_l}} & = & \frac{1}{(\tilde{N_l}\tilde{M})^{j_l}}\left[
   \sum_{\stackrel{l=0}{\omega\not>0}}^{j_l}{j_l \choose l}(-N_l M)^l
(\tilde{N_l}\tilde{M})^{j_l-l}
   \right],\nonumber\\
\hat{K^{j_k}} & = & \frac{1}{(\tilde{N_k}\tilde{M})^{j_k}}\left[
   \sum_{\stackrel{k=0}{\omega\not>0}}^{j_k}{j_k \choose k}(-N_k M)^k
(\tilde{N_k}\tilde{M})^{j_k-l}
   \right],\nonumber\\
\hat{L^{j_l}_0} & = & \frac{1}{(\tilde{N_l}\tilde{M})^{j_l}}\left[
   \sum_{\stackrel{l=1}{\omega\not>0}}^{j_l}{j_l \choose l}(-N_l M)^l
(\tilde{N_l}\tilde{M})^{j_l-l}
   \right],\nonumber\\
\hat{K^{j_k}_0} & = & \frac{1}{(\tilde{N_k}\tilde{M})^{j_k}}\left[
   \sum_{\stackrel{k=1}{\omega\not>0}}^{j_k}{j_k \choose k}(-N_k M)^k
(\tilde{N_k}\tilde{M})^{j_k-l}
   \right],\label{e6}
\eea
where $\sum_{\omega\not>0}$ means summation only over divergent contributions.
We obtain
\bea
\frac{T^{(n_l,n_k)}}{N_l M N_k} & = &
\frac{T^{(n_l,n_k)}}{N_l M N_k} \left[ \hat{L^{j_l}}\hat{K^{j_k}}\right]-
  \frac{T^{(n_l,n_k)}}{N_l M N_k}\left[
  \hat{L^{j_l}} \hat{K^{j_k}}\right]_0.\label{e7}
\eea

Still the first of the two terms on the right hand side of equation (\ref{e7})
can be calculated in
$D=4$ dimensions. We will now consider an example which will be useful
to demonstrate the method.
Consider Fig.(1) with propagators
\beas
P_1 & = & l^2-m_1^2,\\
P_2 & = & (l+q)^2-m_2^2,\\
P_3 & = & (l+k)^2-m_3^2,\\
P_4 & = & (k-q)^2-m_4^2,\\
P_5 & = & k^2-m_5^2,
\eeas
and choose
\beas
T^{(2,1)} & = & l_\mu l_\nu k_\sigma,\\
N_l & = & P_1 P_2,\\
M & = & P_3,\\
N_k & = & P_4 P_5,\\
\tilde{N_l} & = & l^4,\\
\tilde{M} & = & (l+k)^2,\\
\tilde{N_k} & = & k^4,\\
\rightarrow \omega=-1, & & \omega_l=0,\;\;\omega_k=1,\;\; j_l=2,\;\;j_k=0,
\eeas
so that we have
\bea
\frac{T^{(2,1)}}{N_l M N_k}L^2 & = & \frac{T^{(2,1)}}{N_l M
N_k}-2\frac{T^{(2,1)}}{\tilde{N_l} \tilde{M}
          N_k}+\frac{T^{(2,1)}N_l M}{(\tilde{N_l} \tilde{M})^2 N_k},\nonumber\\
\frac{T^{(2,1)}}{N_l M N_k}L^2_0 & = & -2\frac{T^{(2,1)}}{\tilde{N_l} \tilde{M}
          N_k}+\frac{T^{(2,1)}N_l M}{(\tilde{N_l} \tilde{M})^2 N_k},\nonumber\\
\frac{T^{(2,1)}}{N_l M N_k}\hat{L^2} & = & \frac{T^{(2,1)}}{N_l M
N_k}-\frac{T^{(2,1)}}{\tilde{N_l} \tilde{M}
          N_k}+\frac{T^{(2,1)}(2l\cdot q)}{\tilde{N_l} \tilde{M}l^2
N_k},\nonumber\\
\frac{T^{(2,1)}}{N_l M N_k}\hat{L^2_0} & = & -\frac{T^{(2,1)}}{\tilde{N_l}
\tilde{M}
          N_k}+\frac{T^{(2,1)}(2l\cdot q)}{\tilde{N_l} \tilde{M}l^2 N_k},
\label{e8}
\eea
where we used that $P_1=\tilde{P_1}-m_1^2,P_2=\tilde{P_2}+2l\cdot q
+q^2-m_2^2,
P_3=\tilde{P_3}-m_3^2$.  $\tilde{P_i}$ means again the suppression of
all masses and exterior momenta in the corresponding propagator.

We will continue later with this example and discuss now how to calculate
the UV divergent part of equation (\ref{e7}), which is contained in the second
term of the right hand side of this equation.
These integrals are of the form
\bea
\frac{T^{(n_l,n_k)}}{(l^2)^{i}(l+k)^j N_k},\label{e9}\\
\frac{T^{(n_l,n_k)}}{(k^2)^{i}(l+k)^j N_l},\label{e10}\\
\mbox{or}\;\;\frac{T^{(n_l,n_k)}}{(l^2)^{i}(l+k)^j (k^2)^r}\equiv 0.\label{e11}
\eea
The second equation, (\ref{e10}), is just a relabeling of (\ref{e9}).

The third equation, (\ref{e11}), vanishes
identically in dimensional regularization. The integer powers $i,j$ in the
above
equations are determined by $j_l,j_k$.

Let us comment on the vanishing of the integral in (\ref{e11}) here.
It might happen
that this cancellation is due to a mutual cancellation of an
infrared and an ultraviolet divergence, as it appears
in the integral $\int d^D\!l\;/l^4$. Now the UV singular part of the integral
in (\ref{e11})
will have its counterpart in the final integral representation in (\ref{e7}),
where it
is needed to cancel some other UV singular parts. So it seems that we
have installed a spurious infrared singularity in our integral representation.
This should then be regularized by some appropriate IR cutoff and this would
install a corresponding scale in the integral in (\ref{e11}) which would  make
it
non-vanishing.
But these infrared singularities   appear in the integral representation
derived in \cite{dirk} always
as endpoint singularities. As this problem with spurious
IR singularities can only appear when
simultaneously $n_l>0$ and $n_k>0$ it will (in the Feynman gauge) be restricted
to
the
two-loop boson self-energies. Nevertheless it is true in general that
IR singularities appear as endpoint singularities in the method of
\cite{dirk}
which avoids
parametrizations. Let us study these endpoint singularities a little bit more.
Investigating
 the case
\beas
N_l & = & P_1,\\
M & = & P_3,\\
N_k & = & P_5,
\eeas
we find the following expression for the subtraction of UV divergences
\bea
\frac{1}{N_l M N_k}-\frac{1}{\tilde{N_l} \tilde{M} N_k}-\frac{1}{
N_l \tilde{M} \tilde{N_k}}
+\frac{\tilde{M}-m_3^2}{\tilde{N_l} \tilde{M}^2 \tilde{N_k}}.\label{e12a}
\eea
The last term $\sim m_3^2$ is the problematic one.
The integral representation for it can be written as
\bea
\int_{-\infty}^{\infty} dx \int_{-\infty}^{\infty} dy
\frac{1}{\mid x\mid}\frac{1}{\mid x\mid + \mid y\mid + \mid
x-y\mid},\label{e12}
\eea
where we used $\sqrt{x^2+i \eta}=\mid\! x\!\mid$ in the limit
$\eta \to 0$.
Equation (\ref{e12}) has an apparent endpoint singularity at $x=0$.
But these endpoint singularities are a well defined distribution
in the $\eta$-limit when interpreted as Hadamard's Finite Part $HFP$
\cite{Zemanian}.
Its relation
to the propagator is known for long \cite{Nakanishi}.
In fact, the distribution $\int_{-\infty}^{\infty} dx/\sqrt{x^2+i\eta}$
equals  $HFP[1/\mid\! x\!\mid]$. Using this, we handle these
endpoint singularities with the help of $HFP$  and find the corresponding
integral representation
\beas
\int_{1}^{\infty} dx \int_{-\infty}^{\infty} dy
(\frac{1}{\mid x\mid}\frac{1}{\mid x\mid + \mid y\mid + \mid x-y\mid}
 + {x \leftrightarrow -x})
 +\ldots.
\eeas
Here the dots correspond to finite terms for which the endpoint singularity
in $x$
has canceled out after the $y$ integration and so $HFP$ reduces to an ordinary
Riemann integral. As $HFP$ is a local operation which modifies test functions
only
at the location of the singularity our UV behaviour remains unchanged and
the sum in equation (\ref{e12a}) is still UV finite. This can also be
explicitly checked by working out all the different cases for the modulus
function
in equation (\ref{e12}) in detail.
So the integral representation for equation (\ref{e12a}) is  free of IR and UV
singularities.
One can also use $HFP$ to separate nonspurious IR singularities when they
appear.
As for the integral (\ref{e11}) to appear we need UV singular behaviour in
both loop variables,
it is clear that the above subtlety can only arise when there is a
logarithmic UV divergence in the second loop integration, which is exactly
the
case for the above example. In this sense the above example is generic and
serves as a proof in general.

We can still maintain the vanishing of equation (\ref{e11}) by using $HFP$
in our integral representations.
This can also be interpreted as a consistency between $HFP$ and DR.

Let us continue now our general discussion.
We can do one integration in equation (\ref{e9}) leading to remaining
integrals
of the form
\bea
C\int d^Dk \frac{\tilde{T}}{(k^2)^\alpha N_k},\label{e13}
\eea
where $\alpha$ is an integer only when $D=4$ and $\tilde{T}$ is some tensor
in $k$.
The coefficient $C$ will contain a pole in $(D-4)$ in general. This and
possible UV divergences of the $k$ integration itself forbid to calculate
these integrals in four dimensions.

Calculating the above integrals (\ref{e13}) for arbitrary $D$ and $\alpha$ is
possible
but a difficult analytical task. But there is an algebraic approach to these
integrals. We will first transform the UV divergent integrals in equation
(\ref{e7})
via
a partial integration in such a way that the first $l$ integration
is UV finite and its UV divergences are shifted to the $k$ integration.
Then we use a method similar to the method above to split the final
$k$ integration once more into a finite integration, analytically
computable in an easy manner and  UV divergent integrals of massive tadpole
structure. As a consequence everything is expressible by algebraic methods in
one-loop functions.

Defining
\bea
I^{(i,n,m)} & := & \int d^Dl \frac{l_{\mu_1}\ldots l_{\mu_i}}{
    (l^2)^n((l+k)^2)^m}, \;\;\mbox{we have}\nonumber\\
0 & = & \int d^Dl \frac{\partial}{\partial l_\rho}\frac{l_\rho l_{\mu_1}\ldots
l_{\mu_i}}{
    (l^2)^n((l+k)^2)^m}\\
  & = & (D+i-2n-m)I^{(i,n,m)}-mI^{(i,n-1,m+1)}+m k^2 I^{(i,n,m+1)},\nonumber\\
  & & I^{(i,0,m)}=I^{(i,m,0)}=0\;\; \forall i,m.\label{e14}
\eea
As $D+i \geq 2(n+m)$ for UV divergent integrals, we have
$(D+i-2n-m)\not= 0$. Therefore we can always solve equation
(\ref{e14}) for $I^{(i,n,m)}$
in the following.

We can repeat this procedure until we have expressed
$I^{(i,n,m)}$ through integrals of the form $I^{(i,n^\prime,m^\prime)}$
with powers $m^\prime,n^\prime$ such that they are all finite in the
$l$ integration.

We will end up with integrals of the form
\beas
\int d^Dk \frac{k_{\mu_1}\ldots k_{\mu_r}(k^2)^\beta}{N_k},
\eeas
where again $\beta$ is an integer only if $D=4$ but all coefficients of these
integrals are
now finite in four dimensions.

But the $k$ integration is still UV singular, and to find an expression
for these integrals we do a last replacement
\bea
\frac{1}{N_k} & = & \frac{1}{N_k}(\hat{\underline{K}^i}-
\hat{\underline{K}^i_0})\;\;\mbox{where}\label{e15}\\
& & \underline{K}^i:=\frac{(\dot{N_k}-N_k)^i}{\dot{N_k}^i}\;\mbox{and}
\nonumber\\
& & \dot{N_k}:=N_k\mid_{q_i =0},\nonumber
\eea
where it is understood that when expanding the numerators the sums run only
over
the divergent parts, as indicated by the $\hat{{}}$ on the various $K$. Note
that we
have chosen propagators with vanishing
exterior momenta but non-vanishing masses to profit from symmetric
integration properties and we remind the reader that we cannot set the masses
to zero
 because
this would result in vanishing tadpole integrals.
So we end up with finite integrals and separated UV divergent integrals of
the form
\beas
\frac{k_{\mu_1}\ldots k_{\mu_r}(k^2)^\beta}{\dot{N_k}},
\eeas
which are, as they are massive tadpole integrals, easy to evaluate in $D$
dimensions.
The finite integrals in (\ref{e15}) are related to standard massive one-loop
integrals
with integer powers of propagators.

Let us continue our example now. By inspecting equation (\ref{e8}) we see
that we have
to consider $
I^{(2,2,1)}$ and $I^{(3,3,1)}$. The last one is already finite in the $l$
integration. For the first
one we have
\beas
I^{(2,2,1)} & = & \frac{-1}{(D-3)}\left[ \frac{2}{(D-2)} k^2 I^{(2,1,3)} -
  k^2 I^{(2,2,2)}\right].
\eeas
The remaining integrals are of the form
\beas
\int d^Dk \frac{k_{\mu_1}\ldots k_{\mu_i}}{P_4 P_5 (k^2)^{j+\frac{4-D}{2}}}.
\eeas
By using $P_5=\dot{P_5}, P_4=\dot{P_4}-2k\cdot q +q^2$ we separate this
according
to the above equations (\ref{e15}). Note that as a consequence of our
subtraction operation
only terms with an even power of $k$ in the numerator give contributions
to UV divergent integrals. The finite integrals are of the form
\beas
\int d^Dk \frac{k_{\mu_1}\ldots k_{\mu_i}(-2k\cdot q +q^2)^r}{\dot{P_4}^m P_4
P_5 (k^2
)^{j+\frac{4-D}{2}}},
\eeas
where only terms of even power in $k$ contribute in
the numerator.

Via partial fraction decompositions in $k^2,P_5,\dot{P_4}$ this can be
reduced to  standard one-loop
integrals with integer powers of propagators which are algebraically related
to one-loop integrals with unit powers of propagators via mass derivatives.

All these reductions and separations can be handled by symbolic calculation
routines.
So it is possible to reduce an arbitrary two-loop tensor integral into finite
integral representations and a standard set of one-loop integrations.
For the examples discussed in this paper this was always possible
using REDUCE \cite{reduce} on a 486 notebook within less than 5 minutes.
Note that even for the two-loop four-point function $N_k$ is not worse than a
one-loop three-point function (apart from the trivial
topology arising from self-energy insertions), so that the resulting one-loop
integrals are
relatively simple. The knowledge of arbitrary tensor integrals of
one-loop two- and three-point functions with integer powers of propagators is
indeed sufficient for the calculation of two-loop problems by
this method.

Having demonstrated the handling of the UV divergent part let us now discuss
the
integral representations for the finite integrals. In \cite{dirk}
an integral representation is given which transforms the integral to a
twofold integral over the parallel space variables. This can be generalized
to arbitrary tensor integrals which appear as polynomials in parallel
and orthogonal space variables in the numerator. So we have to integrate out
the
orthogonal space variables. This can be easily handled by use of partial
integrations and applying the residue theorem afterwards. But a little
modification
arises for the terms which subtract the UV divergence.  Here we cannot use
a partial fraction decomposition in the $\tilde{P_1},\tilde{P_2}$
propagator as these propagators are equal. Doing the $z$ and
$k_\bot$-integration
as in \cite{dirk}  we are left with
\beas
\int dl_\bot \frac{l_\bot}{(l_0^2-l_\bot^2)^i}[\log
\;\mbox{terms}]; \;\;i=2,3.
\eeas
But we can integrate these terms most easily by rewriting
\bea
\frac{1}{(l_0^2-l_\bot^2)^i} & = & \frac{1}{i!}
\frac{\partial^i}{\partial\mu^i}\frac{1}{(l_0^2-l_\bot^2-\mu)}\mid_{\mu=0}
\label{e16}.
\eea

A similar remark applies to the three-point two-loop functions.
The case of four-point functions will be discussed elsewhere, but will give
no modifications in general \cite{dirk4p}. Nevertheless we include the
principal
reduction of two-loop box integrals in the discussion below.

Up to now we have studied particular tensor integrals only. Let us now discuss
a method
how to calculate tensor integrals more directly. We again use the notation of
\cite{dirk}
for the parallel and orthogonal space variables and will discuss as a specific
example the following graph, which appears in the flavour changing
fermion self-energy. The two-loop radiative corrections to order
${\cal O}(\alpha_s)\times g^2$ are currently being investigated for these
self-energies \cite{dk}.\\[5mm]
%___________________________________________________________________________
% FIGURE
%
\linethickness{0.4pt}
\begin{picture}(33.00,14.00)
\put(16.00,7.00){\circle{14.00}}
\put(23.00,7.00){\line(1,0){10.00}}
\put(9.00,7.00){\line(-1,0){9.00}}
\put(16.00,13.00){\line(0,-1){13.00}}
\put(16.00,13.00){\line(0,1){1.00}}
\put(9.00,14.00){\makebox(0,0)[cc]{1,$q_j$}}
\put(23.00,14.00){\makebox(0,0)[cc]{5,W}}
\put(9.00,0.00){\makebox(0,0)[cc]{2,g}}
\put(23.00,0.00){\makebox(0,0)[cc]{4,$q_i$}}
\put(18.00,7.00){\makebox(0,0)[cc]{3,$q_i$}}
\end{picture}\nopagebreak
%\tenrm\baselineskip=12pt
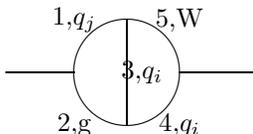
{\small Fig.2: Again the master topology; the labels specify propagators
and particles.\\[5mm]}

Let us decompose the Clifford algebra
\beas
\lslash & = & l_{\mid\mid} \gamma^{\mid\mid} - l_{\bot i} \gamma_\bot^i,\\
\kslash & = & k_{\mid\mid} \gamma^{\mid\mid} - k_{\bot i} \gamma_\bot^i,\\
\{\gamma_{\mid\mid},\gamma_\bot\} & = & 0,
\eeas
where we simply use the fact that we can split a loop momentum in DR in
its component in the direction of the exterior momentum
$q=q_{\mid\mid}$ and its orthogonal
complement. The same splitting can be done for higher dimensional
parallel spaces (the linear span of the exterior momenta). This gives rise to
a corresponding splitting in the Clifford algebra which can also be written
down in a covariant manner
\beas
l_{\mid\mid\mu} & = & \frac{l\cdot q}{q^2}q_\mu,\\
l_{\bot\mu} & = & l_\mu-\frac{l\cdot q}{q^2}q_\mu,\\
\gamma_{\bot\mu} & = & \gamma_\mu-\frac{\qslash}{q^2}q_\mu,\\
\gamma_{\mid\mid\mu} & = &  \frac{\qslash}{q^2}q_\mu,\\
g_{\bot\mu\nu} & = & g_{\mu\nu}-\frac{q_\mu q_\nu}{q^2},\\
\{\gamma_{\mid\mid\nu},\gamma_{\bot\mu}\} & = &
\{\frac{\qslash}{q^2}q_\nu,\gamma_\mu-\frac{\qslash}{q^2}q_\mu\}
=0.
\eeas
The generalization to higher dimensional parallel spaces is clear.

We can now use this decomposition for a reordering
where we bring all parallel space $\gamma$ matrices and the
anticommuting \gf\ \cite{dirkg5} to the left
so that we have
\beas
\gf^{n_5}\gamma_{\mid\mid_1}^{n_1}\ldots\gamma_{\mid\mid_i}^{n_i}
\gamma_{\bot\mu_1}\ldots\gamma_{\bot\mu_{k}},
\eeas
for an $i$ dimensional parallel space and a rank $k$ tensor in the
orthogonal space. The $n_i$ are the number of appearances of \gf\ and
parallel space $\gamma$-matrices.
Using $\gf^2=\one,\gamma_{\mid\mid_i}^2=\one$ we can simplify the above
expession. For the case that we do not have free Lorentz indices in our
Green function we can also simplify the orthogonal space
part of this expression. In this case we
are allowed to replace the orthogonal space $\gamma_\bot$ matrices
identically by metrical tensors
$g_\bot, (g_{\bot\mu}^\mu=(D-i))$, even when we do not
apply a trace to the string. Note that this means that all the lengthy trace
and $\gamma$ algebra calculations
of perturbation theory will not appear for the case of no free Lorentz
indices.
It also means that we only have to consider contributions with an even number
of
$\gamma$-matrices in the orthogonal space.
To prove this statement note that as long as we have no free indices (after
contracting out all double indices) $\gamma_\bot$-matrices can only be
generated by the loop momenta $\lslash,\kslash$. As the only covariant
available for the orthogonal space integration is the total symmetric
combination of metric tensors
$g_\bot$ (the Levi Civita tensor does not appear in symmetric integration)
we only need the symmetric part of the Clifford
algebra in orthogonal space. This corresponds to a replacement of $\gamma_\bot$
matrices by $g_\bot$ tensors, as all indices have to be contracted by
a total symmetric tensor.

This method can be generalized. In the case that we cannot avoid
free indices we still can use the above argument for the $\gamma_\bot$
matrices generated by orthogonal space components of loop momenta.
But note that in most cases it is possible to contract free indices
with metrical tensors or exterior momenta without losing information.
For example, to determine the form factors appearing in radiative corrections
of the fermionic vertex, one can contract with exterior momenta $p,p^\prime$
say. Then one can use all advantages of the case of no free indices.

Our example was calculated using this method by installing the proposed
ordering algorithm
via simple $LET$ rules in REDUCE. Comparing this with the same calculation
using the high energy packet in REDUCE, it was found that the method we
suggest here was orders of magnitude quicker. Even for the one-loop case
we can report on a calculation of two- and three-point functions
in the unitary gauge in the SM where significant advantages were
obtained for the $\gamma$ algebra \cite{Stemler}.

This method fails for the anomaly. The interesting covariant is
$\sim \epsilon_{\mu\nu\rho\sigma}p_1^\rho p_2^\sigma$ and there exists
no tensor constructed from symmetric metrical tensors and exterior
momenta $p_1,p_2$ to give a nonvanishing contraction with this covariant.
One nevertheless can calculate the resulting trace by this method.
But then one has to consider the case of \gf\ with six $\gamma$ matrices,
two in the linear span of $p_1,p_2$, two in the orthogonal complement but
still in four dimensional Minkowski space and two with arbitrary indices,
which gives rise to the non-cyclic trace of \cite{dirkg5}.

Continuing with our example we are left with the tensors
$\one,\gf,\qslash,\gf\qslash$.
Each of these tensors is multiplied by an integral representation
for the finite part, which is of the form (using (\ref{e7}),(\ref{e16})
and the results in \cite{dirk})
\beas
\frac{1}{(P_1-P_2)(P_4-P_5)}\left[
c_{25} {\cal L}_{325}
-
c_{15} {\cal L}_{315}
+
c_{14} {\cal L}_{314}
-
c_{24} {\cal L}_{324}\right]\\
-
\frac{\partial}{\partial\mu}
\left[
(c(3,0)+c(2,1)+c(2,0)+c(1,1)){\cal L}_{\mbox{\tiny sub}}
\right]
\!{}_{\mid_{\mu=0}}
- l\cdot q
\frac{\partial^2}{\partial\mu^2}
\left[
(c(3,0)+c(2,1)){\cal L}_{\mbox{\tiny sub}}
\right]
\!{}_{\mid_{\mu=0}}\\
-
\frac{\partial}{\partial\mu}
\left[
(c(0,3)+c(1,2)+c(0,2)){\cal K}_{\mbox{\tiny sub}}
\right]
\!{}_{\mid_{\mu=0}}
+k\cdot q
\frac{\partial^2}{\partial\mu^2}
\left[
(c(0,3)+c(1,2)){\cal K}_{\mbox{\tiny sub}}
\right]
\!{}_{\mid_{\mu=0}}.
\eeas
We list the explicit coefficients $c_{25},c_{15},c_{24},c_{14},c(i,j)$ in an
appendix.
For the case of Fig.(2) it took only about 20 seconds in REDUCE to
generate them
for the complete graph.

Let us summarize the situation for the most interesting
two-loop graphs. We assume that the corresponding calculations
are done in a renormalizable gauge, though it is
not necessary to  restrict to this case.\\[5mm]
{\large Two-point functions:}\\
We have two topologies.
For the trivial self-energy insertions we can always arrange
that power counting improves by  $2n_l$ (assuming
$l$ is the subloop momentum), as we can always arrange exterior momenta not
to flow through the subloop. For this topology we then have
$n_k=0$ always, while $n_l=1$ or $n_l=2$ corresponding to $\omega=1$
or
$\omega=2$, that is corresponding to boson or fermion self-energies.

For the other (master) topology of our example we have $(n_k=0, n_l=2)$
or vice versa if $\omega=1$ and $(n_l=2,n_k=1)$ or vice versa if
$\omega=2$.\\[5mm]
{\large Three-point functions:}\\
We have three topologies, the ladder topology, the crossed topology
and the self-energy insertion topology. For the last one we always have
$(n_l=1,n_k=0)$ while the other ones reduce to $(n_k=0, n_l=1)$
or $(n_k=0, n_l=2)$, depending what kind of vertex one considers
(the fermionic
one has only logarithmic overall divergence,
while e.g.~triple boson couplings are
linear divergent). Note that for the crossed topology, as $M$ contains
two propagators, the UV divergent one-loop integrals reduce to two-point
functions and so this more complicated topology is simpler in
its divergent structure. This corresponds to the fact that it has no proper
subdivergence, while the ladder topology has a vertex type subdivergence and
thus
involves three-point one-loop functions.\\[5mm]
{\large Four-point functions:}\\
We need four exterior bosons to have a UV divergence of logarithmic degree.
We then have $\omega=0$ and  the case $(n_l=1,n_k=0)$ is always
sufficient. Again the complicated topologies involve simpler
one-loop integrals.\\[5mm]

We see that only for boson self-energies we have the case $(n_l\not= 0,
n_k\not= 0)$ so that only for them do we have to use $HFP$
explicitly. Note also that all $(n_l=1,n_k=0)$  cases
reduce to a very simple subtraction which simply sets the exterior momenta
and masses to zero in the subtracted term, and this is sufficient
especially for the two-loop fermionic vertex corrections.

We conclude that an easy separation of divergences for two-loop functions
can be achieved. We saw that for the subtraction of the UV divergences
of the real two-loop
part very simple one-loop functions were sufficient. Especially
only terms of order $1/(D-4)$ were used. This was not unexpected as
the order $1/(D-4)^2$ poles of a two-loop graph can be obtained from
products of one-loop functions. This is true in general for the
leading divergence in a $n$-loop calculation. So we will find the leading
divergences in the terms where the $M$ part of the denominator is cancelled,
giving rise to decoupled one-loop integrals. For our example in
Fig.(2) we list the corresponding expressions in the appendix.
As the individual terms of equation (\ref{e7}) correspond
to terms subtracting $l$ divergences, $k$ divergences and
overall divergences, one might wonder if it is possible to implement
the whole renormalization program (that is the addition of
one-loop counterterm graphs) at this level to obtain even simpler expressions.
We intend to investigate this question in the future.\\[1cm]
{\large Acknowledgements\\}
It is a pleasure to thank D.~Broadhurst and K.~Schilcher for stimulating
discussions on the subject. This work was supported in part by the
Deut\-sche For\-schungs\-ge\-mein\-schaft and under grant number
A69231484 from the Australian Research Council.\\[1cm]
{\Large Appendix\\[5mm]}
\begin{appendix}
In this appendix we list the relevant polynomials
in $l_0,k_0$ for the final integral representation
of the graph of Fig.(2) explicitly. Our notation follows
\cite{dirk}. We list these results to give an idea of the complexity
resp.~simplicity of the integral representation for a complete
graph. We see that the integral representation, compared with
the pure integral representation for the scalar master function,
is similar in its qualitative and quantitative difficulty.
Let us first list the propagators of Fig.(2):
\beas
P_1  :=  l^2-m_j^2 & & P_2 := (l-q)^2,\\
P_3 := (l+k)^2-m_i^2, & &\\
P_4 := (k+q)^2-m_i^2 & & P_5 := k^2-m_w^2.
\eeas
We define $q:=\sqrt{q^2}$ and $\hat{\qslash}:=\qslash/q$.
$$\displaylines{\qdd
c_{15}
:=
\[8\cdot \gf\hat{\qslash}
  \cdot
  \(-
    \(2\cdot q\cdot l_{0}^{2}
    \)
    \nl
    -2\cdot q\cdot l_{0}\cdot k_{0}
    +m_w^{2}\cdot l_{0}
    +m_w^{2}\cdot k_{0}
    -m_i^{2}\cdot k_{0}
    +m_j^{2}\cdot l_{0}
    +m_j^{2}\cdot k_{0}
  \)
  \nl
  +4\cdot \gf
  \cdot m_j\cdot
  \(-
    \(2\cdot q\cdot l_{0}
    \)
    -2\cdot q\cdot k_{0}
    -m_w^{2}
    +3\cdot m_i^{2}
    +m_j^{2}
  \)
  \nl
  +8\cdot \hat{\qslash}
  \cdot
  \(-
    \(2\cdot q\cdot l_{0}^{2}
    \)
    -2\cdot q\cdot l_{0}\cdot k_{0}
    +m_w^{2}\cdot l_{0}
    +m_w^{2}\cdot k_{0}
    -m_i^{2}\cdot k_{0}\nl
    +m_j^{2}\cdot l_{0}
    +m_j^{2}\cdot k_{0}
  \)
  +4\cdot m_j\cdot
  \(2\cdot q\cdot l_{0}
    +2\cdot q\cdot k_{0}
    +m_w^{2}
    -3\cdot m_i^{2}
    -m_j^{2}
  \)
\]
\nl}$$
$$\displaylines{\qdd
c_{14}
:=
\[8\cdot \gf\hat{\qslash}
  \cdot
  \(-
    \(q^{2}\cdot l_{0}
    \)
    \nl
    -q^{2}\cdot k_{0}
    -2\cdot q\cdot l_{0}^{2}
    -4\cdot q\cdot l_{0}\cdot k_{0}
    -2\cdot q\cdot k_{0}^{2}
    +m_i^{2}\cdot l_{0}
    +m_j^{2}\cdot l_{0}\nl
    +m_j^{2}\cdot k_{0}
  \)
  +4\cdot \gf
  \cdot m_j\cdot
  \(q^{2}
    -2\cdot q\cdot l_{0}
    +2\cdot m_i^{2}
    +m_j^{2}
  \)
  \nl
  +8\cdot \hat{\qslash}
  \cdot
  \(-
    \(q^{2}\cdot l_{0}
    \)
    -q^{2}\cdot k_{0}
    -2\cdot q\cdot l_{0}^{2}
    -4\cdot q\cdot l_{0}\cdot k_{0}
    -2\cdot q\cdot k_{0}^{2}\nl
    +m_i^{2}\cdot l_{0}
    +m_j^{2}\cdot l_{0}
    +m_j^{2}\cdot k_{0}
  \)
  +4\cdot m_j\cdot
  \(-q^{2}
    +2\cdot q\cdot l_{0}
    -2\cdot m_i^{2}
    -m_j^{2}
  \)
\]
\nl}$$
$$\displaylines{\qdd
c_{25}
:=
\[8\cdot \gf\hat{\qslash}
  \cdot
  \(-
    \(q^{2}\cdot l_{0}
    \)
    -q^{2}\cdot k_{0}\nl
    +m_w^{2}\cdot l_{0}
    +m_w^{2}\cdot k_{0}
    -m_i^{2}\cdot k_{0}
  \)
  +4\cdot \gf
  \cdot m_j\cdot
  \(-q^{2}
    -2\cdot q\cdot k_{0}
    -m_w^{2}\nl
    +3\cdot m_i^{2}
  \)
  +8\cdot \hat{\qslash}
  \cdot
  \(-
    \(q^{2}\cdot l_{0}
    \)
    -q^{2}\cdot k_{0}
    +m_w^{2}\cdot l_{0}\nl
    +m_w^{2}\cdot k_{0}
    -m_i^{2}\cdot k_{0}
  \)
  +4\cdot m_j\cdot
  \(q^{2}
    +2\cdot q\cdot k_{0}
    +m_w^{2}
    -3\cdot m_i^{2}
  \)
\]
\nl}$$
$$\displaylines{\qdd
c_{24}
:=
\[8\cdot \gf\hat{\qslash}
  \cdot
  \(-
    \(2\cdot q^{2}\cdot l_{0}
    \)
    \nl
    -2\cdot q^{2}\cdot k_{0}
    -2\cdot q\cdot l_{0}\cdot k_{0}
    -2\cdot q\cdot k_{0}^{2}
    +m_i^{2}\cdot l_{0}
  \)
  +8\cdot m_i^{2}\cdot m_j\cdot \gf
  \nl
  +8\cdot \hat{\qslash}
  \cdot
  \(-
    \(2\cdot q^{2}\cdot l_{0}
    \)
    -2\cdot q^{2}\cdot k_{0}
    -2\cdot q\cdot l_{0}\cdot k_{0}
    -2\cdot q\cdot k_{0}^{2}
    +m_i^{2}\cdot l_{0}
  \)
  -8\cdot m_i^{2}\cdot m_j
\]
\nl}$$
$$\displaylines{\qdd
c(3,0):=
\[8\cdot l_{0}\cdot \mu\cdot \gf\hat{\qslash}
  +8\cdot l_{0}\cdot \mu\cdot \hat{\qslash}
\]
\cr}$$
$$\displaylines{\qdd
c(2,1):=
\[8\cdot k_{0}\cdot \mu\cdot \gf\hat{\qslash}
  +8\cdot k_{0}\cdot \mu\cdot \hat{\qslash}
\]
\cr}$$
$$\displaylines{\qdd
c(1,2):=
\[8\cdot l_{0}\cdot \mu\cdot \gf\hat{\qslash}
  +8\cdot l_{0}\cdot \mu\cdot \hat{\qslash}
\]
\cr}$$
$$\displaylines{\qdd
c(0,3):=
\[8\cdot k_{0}\cdot \mu\cdot \gf\hat{\qslash}
  +8\cdot k_{0}\cdot \mu\cdot \hat{\qslash}
\]
\cr}$$
$$\displaylines{\qdd
c(2,0):=
\[-
  \(16\cdot q\cdot l_{0}^{2}\cdot \gf\hat{\qslash}
  \)
  +4\cdot m_j\cdot \mu\cdot \gf
  -16\cdot q\cdot l_{0}^{2}\cdot \hat{\qslash}
  -4\cdot m_j\cdot \mu
\]
\nl}$$
$$\displaylines{\qdd
c(1,1):=
\[-
  \(16\cdot q\cdot l_{0}\cdot k_{0}\cdot \gf\hat{\qslash}
  \)
  -16\cdot q\cdot l_{0}\cdot k_{0}\cdot \hat{\qslash}
\]
\cr}$$
$$\displaylines{\qdd
c(0,2):=
\[-
  \(4\cdot m_j\cdot \mu\cdot \gf
  \)
  +4\cdot m_j\cdot \mu
\]
\cr}$$
$$\displaylines{\qdd
{\cal L}_{\mbox{sub}}:=
\[
  \(\ln
    \(\sqrt{q^{2}
            +2\cdot q\cdot k_{0}
            -m_i^{2}
            +k_{0}^{2}}
      +
      \sqrt{l_{0}^{2}
            -\mu}
      +\mid\!l_{0}
      +k_{0}\!\mid
    \)
    \nl
    -\ln
    \(\sqrt{l_{0}^{2}
            -\mu}
      +
      \sqrt{
            -m_w^{2}
            +k_{0}^{2}}
      +\mid\!l_{0}
      +k_{0}\!\mid
    \)
  \)
\]/(P_4-P_5)
\cr}$$
$$\displaylines{\qdd
{\cal K}_{\mbox{sub}}:=
\[
  \(-\ln
    \(\sqrt{k_{0}^{2}
            -\mu}
      +
      \sqrt{
            -m_j^{2}
            +l_{0}^{2}}
      +\mid\!l_{0}
      +k_{0}\!\mid
    \)
    \nl
    +\ln
    \(\sqrt{k_{0}^{2}
            -\mu}
      +\mid\! l_{0} - q \!\mid
      +\mid\! k_{0}+l_{0}\!\mid
    \)
  \)
\]/(P_1-P_2)
\cr}$$
$$\displaylines{\qdd
O\!O:=
\[-
  \(8\cdot \gf\hat{\qslash}
    \cdot
    \(l_{0}
      +k_{0}
    \)
  \)
  -4\cdot m_j\cdot \gf
  -8\cdot \hat{\qslash}
  \cdot
  \(l_{0}
    +k_{0}
  \)
  +4\cdot m_j
\]
\nl}$$
where we used the symbol $O\!O$ to denote the term where the $M$ part
of the denominator is cancelled, so that is corresponds to a squared one-loop
graph.

\end{appendix}

\end{document}